\documentclass[sigconf]{acmart}
\AtBeginDocument{%
  \providecommand\BibTeX{{%
    \normalfont B\kern-0.5em{\scshape i\kern-0.25em b}\kern-0.8em\TeX}}}

\acmYear{2023}\copyrightyear{2023}
\setcopyright{rightsretained}
\acmConference[MMSys '23]{Proceedings of the 14th ACM Multimedia Systems Conference}{June 7--10, 2023}{Vancouver, BC, Canada}
\acmBooktitle{Proceedings of the 14th ACM Multimedia Systems Conference (MMSys '23), June 7--10, 2023, Vancouver, BC, Canada}
\acmPrice{}
\acmDOI{10.1145/3587819.3590976}
\acmISBN{979-8-4007-0148-1/23/06}

\usepackage{sansmath}  
\usepackage[nolist]{acronym}
\usepackage{subfigure}
\usepackage[algo2e]{algorithm2e} 
\usepackage{algorithmic}
\usepackage{hyperref}
\usepackage{booktabs}
\usepackage{adjustbox}
\usepackage{gensymb}
\usepackage{amsmath}
\usepackage{textcomp}
\usepackage{bm}
\usepackage{xcolor}
\usepackage{multirow}
\usepackage{multicol}
\usepackage{soul}
\usepackage{blindtext}

\newcommand{\ie}[0]{\textit{i.e.},~}
\newcommand{\eg}[0]{\textit{e.g.},~}

\newcommand{\sdof}[0]{6-\ac{DOF}~}
\newcommand{\ssdof}[0]{6-\ac{DOF}}
\newcommand{\tdof}[0]{3-\ac{DOF}~}

\definecolor{set1}{rgb}{0.325,    0.678,    0.196}
\definecolor{set2}{rgb}{0.9176,    0.4667,    0.1216}
\definecolor{set3}{rgb}{0.1176,    0.5686,    0.9176}
\definecolor{set4}{rgb}{0.1176,    0.9137,    0.4627}

\begin{document}

\title[Extending 3-DoF metrics to model user behaviour  similarity   in 6-DoF immersive applications]{Extending 3-DoF metrics to model user behaviour \\ similarity in 6-DoF immersive applications}

\author{Silvia Rossi}
\affiliation{%
 \institution{Centrum Wiskunde \& Informatica (CWI)}
 \city{Amsterdam}
 \country{the Netherlands}
}
\email{s.rossi@cwi.nl}

\author{Irene Viola}
\affiliation{%
 \institution{Centrum Wiskunde \& Informatica (CWI)}
 \city{Amsterdam}
 \country{the Netherlands}
}
\email{irene@cwi.nl}

\author{Laura Toni}
\affiliation{%
  \institution{University College London (UCL)}
  \city{London}
  \country{UK}}
\email{l.toni@ucl.ac.uk}

\author{Pablo Cesar}
\affiliation{%
\institution{CWI, TU Delft}
 \city{Amsterdam}
 \state{the Netherlands}\\
 \city{Delft}
 \country{the Netherlands}
}
\email{p.s.cesar@cwi.nl}
\thanks{This work has been carried out with the support of ERCIM ‘Alain Bensoussan’ Fellowship Programme.}
\renewcommand{\shortauthors}{Rossi, et al.}

\begin{abstract}
Immersive reality technologies, such as Virtual and Augmented Reality, have ushered a new era of user-centric systems, in which every aspect of the coding--delivery--rendering chain is tailored to the interaction of the users. Understanding the actual interactivity and behaviour of the users is still an open challenge and a key step to enabling such a user-centric system. Our main goal is to extend the applicability of existing behavioural methodologies for studying user navigation in the case of  6 Degree-of-Freedom (DoF). Specifically, we first compare the navigation in  6-DoF with its 3-DoF counterpart highlighting the main differences and novelties. Then, we define new metrics aimed at better modelling behavioural similarities between users in a 6-DoF system. We validate and test our solutions on real navigation paths of users interacting with dynamic volumetric media in 6-DoF Virtual Reality conditions. Our results show that metrics that consider both user position and viewing direction better perform in detecting user similarity while navigating in a 6-DoF system. Having easy-to-use but robust metrics that underpin multiple tools and answer the question ``how do we detect if two users look at the same content?" open the gate to new solutions for a user-centric system.
\end{abstract}



\maketitle
	\begin{figure}[b]
		\begin{acronym}
			\acro{AR}{Augmented Reality}
			
			\acro{CAVE}{Cave Automatic Virtual Environment}	
			\acro{CDN}{Content Delivery Network}	
			\acro{CMP}{Cube Map Projection}
			\acro{CNN}{Convolutional Neural Network}
			\acro{CWI}{Centrum Wiskunde \& Informatica}
			
			\acro{DASH}{Dynamic Adaptive Streaming over HTTP}
			\acro{DoF}{Degree of Freedom}
			\acro{DSIS}{Double-Stimulus Impairment Scale}
			\acro{DOF}[DoF]{Degrees-of-Freedom}
			
			\acro{ERP}{Equirectangular Projection}
			
			\acro{FoA}{focus of attention}
			\acro{FoV}{field of view}
			\acro{FP}{False Positive}
			\acro{FPR}{False Positive Rate}
			\acro{FSM}{fused saliency maps}
			
			\acro{GBVS}{Graph-Based Visual Saliency}
			\acro{GDI}{Generalised Dunn's Index}
			
			\acro{HAS}{HTTP adaptive streaming}
			\acro{HQ}{High Quality}
			\acro{HMD}{head-mounted display}
			
			\acro{ILP}{integer linear programming}
			
			\acro{MCP}{maximum clique problem}
			\acro{MWCP}{maximum-weight clique problem}
			\acro{MCTS}{motion-constrained tile sets}
			\acro{MPEG}{Moving Picture Experts Group}
			\acro{MSE}{Mean Square Error}
			
			\acro{OMAF}{omnidirectional media application format}
			
			\acro{PSNR}{peak signal-to-noise ratio}
			\acro{PC}{Point Cloud}
			
			\acro{QEC}{Quality Emphasis Center}
			\acro{QER}{Quality Emphasised Region}
			\acro{QoE}{Quality of Experience}	
			\acro{QoS}{quality-of-service } 
			
			\acro{ROC}{Receiver Operating Characteristic}
			\acro{RoI}{Region of Interest}
			
			\acro{SI}{Spatial Information}
			\acro{SRD}{Spatial Relationship Description}
			
			\acro{VoD}{Video on Demand}
			\acro{VR}{Virtual Reality}
			
			\acro{TCD}{Trinity College of Dublin}
			\acro{TI}{Temporal Information}
			\acro{TP}{True Positive}
			\acro{TPR}{True Positive Rate}
			\acro{TSP}{Truncated Square Pyramid}
		\end{acronym}
	\end{figure}

\vspace{-0.1cm}
\section{Introduction}
\label{sec:intro}
Immersive reality technology has revolutionised how users engage and interact with media content, going beyond the passive paradigm of traditional video technology, and offering more degrees of presence and interaction in a virtual environment. 
Depending on how much a user can move in the 3D space, immersive environments can be classified as 3- or 6-\ac{DOF}. In a \tdof scenario, the de-facto multimedia content is the  \emph{omnidirectional} or \emph{spherical video}, representing an entire $360^\circ$ environment on a virtual sphere. The viewer is fully immersed in a virtual space where they can navigate and interact thanks to an immersive device -- typically a \ac{HMD},  which enables to view only a portion of the environment around themself, named \textit{viewport}. 
The media is displayed from an \emph{inward} position, and the viewer can interact with the content only by changing the viewing direction (\ie by looking up/down or left/right or tilting the head side to side). 
In a \sdof system, the user can also change viewing perspective by moving (e.g., walking,  jumping) inside the virtual space. The scene is therefore populated by \emph{volumetric objects} (\ie meshes or point clouds) which are observed from an \emph{outward} position. 
These extra degrees of freedom bring the virtual experience even closer to reality: a higher level of interactivity makes the user more immersed and present within the virtual environment \cite{cipresso2018past}. 

Despite their differences, the common denominator of both interactive systems 
is the viewer as an active decision-maker of the displayed content. This active role defines the \textit{user-centric} era, in which content processing, streaming, and rendering need to be tailored to the viewer interaction to remain bandwidth-tolerant whilst meeting quality and latency criteria \cite{rossi2022streaming,viola2022volumetric}. Media codecs need to be optimised such that the quality experienced by the user is maximised \cite{xu2020state,schwarz2018emerging}. Similarly, streaming should be tailored to users' interactivity to ensure high-quality content and smooth navigation, while remaining bandwidth-tolerant~\cite{harth2018different,subramanyam2020user,park2019rate}. From here, the importance to understand, analyse and predict  users' movements (\ie \textit{user behaviour}) within an immersive scenario~\cite{rossi2020users,van2019towards,han2020vivo,rondon2021track}. 
A better understanding of how the population behave when experiencing immersive reality has an impact that goes beyond system applications,  leading to user similarities, \ie \textit{user clustering/profiling} \cite{pfeuffer2019behavioural}, which is essential for several purposes: from secure authentication \cite{tricomi2022you} to medical application \cite{martin2021use}.

Thanks to the large availability of public datasets \cite{nasrabadi2019taxonomy,rondon2020unified,jin2022you}, user navigation in \tdof immersive systems has been deeply investigated \cite{sitzmann2018,rossi2020understanding}, showing the  importance of analysing and detecting 
key behavioural aspects  in  interactive  (user-centric) systems.  However, the \sdof counterpart has been scarcely considered in the literature~\cite{alexiou2020pointxr,subramanyam2020comparing,zerman2021user}. 
Due to the change in the viewing paradigm (from inward to outward) and to more level of interaction in 6-DoF, current studies in \tdof cannot be directly applied to \sdof domains ~\cite{rossi2021anewchallenge}. Filling this gap is the main goal of this paper by providing new metrics for user analysis in \ssdof.  

In this work, we focus on extending the applicability of clustering methods to  investigate users similarity (\ie  users sharing common behaviours while interacting with the content) to  \sdof environments. Specifically, clustering techniques usually rely on pairwise similarity metrics, with similarity being in this case in terms of 6-DoF interaction. To the best of our knowledge, such metric has not been proposed yet in  6-DoF context.  Starting from the state-of-the-art clustering algorithm developed in \tdof \cite{rossi2019spherical}, and the main limitations of the tool  when extended to 6-DoF described in \cite{rossi2021anewchallenge}, we investigate new methodologies for better modelling user similarities and overcoming those limitations. 
First, we recall the definition of user navigation trajectory in  6-DoF. 
Then,  we present the  exact user similarity metric, which we will be considering as our ground truth. Given its computational complexity, after an exhaustive study, we propose a simpler and  yet reliable proxy for it.   More concretely, we define and compare 8  similarity metrics which are based on different \textit{distance features} (\ie user positions in the 3D space, user viewing directions) and \textit{distance measurements} 
(\ie Euclidean, Geodesic distance). We validate and test our proposed similarity metrics on a publicly available dataset of navigation trajectories collected in a \sdof \ac{VR} scenario \cite{subramanyam2020user}. Results have shown that similarity metrics based on more distance features are promising solutions to correctly detect users with similar behaviour while experiencing volumetric content. 

Our work contributes to the overall open problem of behavioural analysis in a \sdof system with the following main contributions:
\begin{itemize}
    \item presenting the general problem of detecting behavioural similarities in a \sdof system, and introducing novel similarity metrics able to model the user behaviour in this scenario. These are expressed as a function of various distance features and measurements and we divide them into two groups: \textit{single-} and \textit{multi-features metrics};
    \item an exhaustive analysis of the different proposed metrics aimed at capturing users' trajectory similarity (in terms of distance on the plane or from the object)  and the ability to approximate the  ground truth. This analysis based on 6-DoF VR trajectories reveals that the position on the floor alone is not sufficient to characterise the user behaviour and that the viewing direction cannot be neglected.
\end{itemize}

The remainder of this article is organised as follows: related work on user behavioural analysis in both 3-DoF and \sdof systems are reported in Section~\ref{sec_relatedWork}. The main challenges of detecting behavioural similarities in a \sdof system and the importance of having a tool that approximates such similarities are described in Section~\ref{sec_challenges_new}. Our proposed similarity metrics are described in Section~\ref{sec_proposed_metrics}; while Section~\ref{sec_validation} and Section~\ref{sec_analysis_results} present experimental setup and validation of our proposed metrics on real navigation trajectories collected in a \sdof \ac{VR} setting, respectively. 
Further discussion and final conclusion are summarised in Section~\ref{sec_conclusion}.


\section{Related Work}
\label{sec_relatedWork}
We now describe how user behaviour has been analysed in \tdof systems, showing also the benefit of this type of analysis in user-centric systems.  Then, we show which methods have been used for the analysis in \sdof scenarios, highlighting the still outstanding open challenges. 
\subsection{User Behaviour in 3-DoF environment}
The user navigation within a \tdof environment has been intensely analysed from many perspectives. Many studies have focused on psychological investigations of user engagement and presence correlated to movements within the spherical content.  In \cite{hanseul2020stimulus}, a study from a large-scale experiment (511 users and 80 omnidirectional videos) showed a positive correlation between lower interactivity level and higher engagement level (strong focus on  few  points of interest). 
Similarly, a  correlation between the perceived sense of presence and the interactivity level was detected in~\cite{bermejo2021relationships}, with more random exploratory interactions for less immersed (and hence less engaged)  users. However, no objective metric to properly quantify and characterise user behaviour has been presented in these works.

To further understand how people observe and explore 360\degree~contents, many public datasets of navigation trajectories have been made available. Those datasets usually come with statistical analysis aimed at capturing average users behaviour, as a function of maximum and average angular speeds under various video segment lengths~\cite{corbillon2017}, exploration time \cite{sitzmann2018} or  eye fixation distribution~\cite{david2018dataset}. 
A deeper analysis was presented in \cite{nasrabadi2019taxonomy} where the dataset 
has been analysed through a clustering algorithm presented in~\cite{rossi2019spherical}, specifically built to have in the same cluster users who similarly explore 360\degree~content. 
However, behavioural analysis based on such clustering tool mainly provides a general idea of similarity among viewers without offering however a quantitative metric. To overcome such limitation, authors in \cite{rossi2020understanding}, showed the benefit of studying spatio-temporal trajectories by information theory metrics, and thus the possibility of identifying and quantifying behavioural aspects. Key outcomes from this quantitative analysis were the study of similarities between users when watching the same content, but also the similarity of a given user when watching diverse content.      The importance of these behavioural insights has been then exploited in different \ac{VR} applications. 
For instance, authors in \cite{nasrabadi2020viewport} proposed a scalable prediction algorithm for user navigation, which considered previous navigation patterns while in \cite{Morais2021contentbased} a hybrid approach has been presented based on both dominant user behaviour (detected via a clustering approach) and the video content. Recently, authors in \cite{guimard2022deep} showed that behavioural uncertainty could lead to different navigation in the future even if previously presented similarity; thus, a deep variational learning framework to predict multiple plausible head trajectories was presented. Moreover, in order to extend publicly available navigation datasets, realistic synthetic head rotation data were also generated using a deep learning algorithm given similar data distribution over time \cite{struye2022generating}.
Finally, the analysis and understanding of user navigation in a \ac{VR} environment have shown promising results also in determining the mental health issues of subjects (\eg anxiety, autism spectrum disorder, eating disorders, depression) and their treatment \cite{freeman2017virtual,geraets2021advances,mazumdar2021early}. 
\vspace{-0.1cm}
\subsection{User Behaviour in 6-DoF environment}
Extending such behavioural  analysis to a \sdof environment is not straightforward, due to the change in the viewing paradigm (from inward to outward) and the addition of translation in 3D space. In the past, user navigation in \sdof scenarios was studied in the context of locomotion and display technology for CAVE environments \cite{swindells2004comparing,ragan2016amplified}. A \ac{CAVE} system is an immersive room on which walls and floor are projected the video content  and viewers are free to move inside \cite{Creagh2003}. For instance, the study in \cite{swindells2004comparing} focused on task performance analysis in terms of completion time and correct actions.  
Authors in \cite{ragan2016amplified}  compared instead the effect of two different immersive platforms such as CAVE and \ac{HMD} on the user navigation. 
More traditional metrics, such as angular distance and linear velocity, alongside completion time, were also used to compare different navigation controllers (\ie joystick-based vs head-controlled navigation) in \sdof \cite{chen20136dof}. In detail,  the authors showed the superiority of head-controlled techniques, allowing more sense of presence and better control with less discomfort in the navigation. While the tools mentioned above are highly informative to summarise the interaction of users within a \sdof environment, they usually fail to provide other key insights: which users navigate similarly, and which are the dominant interaction  behaviour among users.

 Recently, the focus has been put on subjective quality assessment based on different coding techniques of volumetric content, both static \cite{alexiou2020pointxr} and dynamic \cite{subramanyam2022subjective}. These studies present a statistical analysis of user movements in terms of mean angular velocity, the ratio of frames viewed while in movement, most displayed areas of the content showing an influence in the navigation due to the perceived content quality, and point out a users' preference to visualise the volumetric object from a close and frontal perspective. 
 A behavioural analysis of user navigating in \sdof social VR movie has been also presented in \cite{rossi2021Influence}. An investigation on how users are affected by virtual characters and narrative elements of the movie has been conducted through objective metrics, showing a more static behaviour when an interactive task was requested, and more exploratory movements during dialogues. Authors in \cite{rossi2022behavioural} present an exploratory behavioural analysis of users while displaying volumetric content within a \sdof environment focusing on the understanding how the way of navigating is affected by the content and its features, such as dynamics and quality, but also by the intrinsic disposition of the single user. Finally, to encourage the collection of navigation data in \sdof immersive experience, a new tool was recently released in \cite{villenave2022xrecho}.
 
 These preliminary studies are based on conventional metrics, which consider only one user feature at a time, either position on the floor or viewing direction but not together, suffering from the major shortcomings highlighted before. In this paper, we aim to overcome these limitations by proposing a general and efficient tool for detecting similar viewers while experiencing \sdof content.

 \begin{figure*}[!ht]
	\centering
    \includegraphics[width=\textwidth]{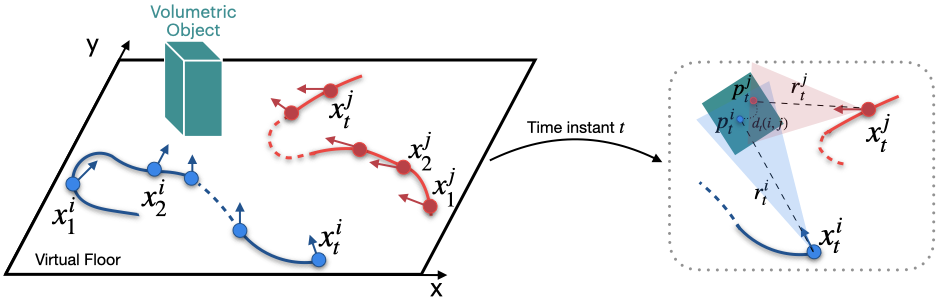} 
	\vspace{-2.5em}
	\caption{Example of two \sdof~trajectories projected in a 2D domain for user $i$ and $j$. On the right side, a snapshot at time $t$: coloured triangles represent viewing frustum per user.}
	\vspace{-1.2em}
	\label{fig_6dofnavigation}
\end{figure*}
\section{Challenges}
\label{sec_challenges_new}
In this work, our main goal is to define a new pairwise metric able to capture the (dis)similarity between two \sdof users (in terms of displayed content). This metric needs to be reliable and yet simple to compute. In the following,  we first present our assumption of similarity among users while navigating in a \sdof environment based on \cite{rossi2021anewchallenge}. Then, we show an exact user similarity metric highlighting its limitations, and therefore the need to find a simpler proxy for it. Finally, we emphasise the advantages of having a similarity metric for behavioural analysis via a clique-based clustering approach presented in \cite{rossi2019spherical}, which  identified users who are attending the same portion of an omnidirectional content in a 3-\ac{DOF} system. This clustering technique relies on a pairwise similarity metric, and thus, having a proper metric also for \sdof system would extend the applicability of this state-of-the-art tool. 

\subsection{User Similarity in 6-DoF}

Similarly to \cite{rossi2021anewchallenge}, we are interested in analysing user behaviour, assuming that users interact similarly when they \emph{observe the same volumetric content.} The user behaviour can be identified by the  spatio-temporal sequences of their movements within the virtual environment, namely \emph{navigation trajectories}.\\

In a \tdof~scenario, the trajectory of a generic user $i$ can be formally denoted by the sequence of the user's viewing direction over time $ \{p^i_1,p^i_2,..,p^i_n\}$ where $p^i_t$ is the centre of the viewport projected on the immersive content (\ie spherical video) at timestamp $t$. 
In fact, the viewport centre alone is highly informative of the user behaviour and can be used as a proxy of viewport overlap among users~\cite{rossi2019spherical}. In particular, the geodesic distance has been proved as a reliable similarity metric such that a low value indicates high similarity between \tdof users. 

Differently in a \sdof setting, the more degrees of freedom are given to the user, the more challenging becomes the system and the description of user navigation within it. The viewport centre alone is no more sufficient to characterise the user behaviour in a \sdof scenario since now the distance between the user and immersive content can change over time due to the added degrees of freedom. Figure~\ref{fig_6dofnavigation} shows an example of two users navigating in a \sdof system. On the left side of the figure, there are navigation trajectories of two users $i$ and $j$ projected on a 2-D domain (\ie floor). Each point $x_t$ represents the spatial coordinates (\ie [x,y,z]) on the floor of viewers while each associated vector symbolises the viewing direction.  In the right part of Figure~\ref{fig_6dofnavigation}, we have instead a snapshot of a specific time instant $t$. In more detail, the shaded triangular areas represent the \emph{viewing frustum} per user, which indicates the region within the user viewport, and $r_t$ is the distance between the user and the volumetric content. We have also depicted the viewport centre $p_t$ projected on the displayed volumetric object. Given the two users $i$ and $j$ at time $t$, in the case of $r^i_t \gg r^j_t$, the user $j$ (very close to the object) is visualising a very focused and detailed part of it; conversely, user $i$ is pointing to the same area but from a much further distance, thus the experienced content is different with less defined details. Despite this difference, the small distance $D_t(i,j)$ between viewport centres $p_t^i$ and $p_t^j$ might suggest a high similarity between the users, which does not reflect the reality in the case of $r^i_t \gg r^j_t$. Thus in this scenario, we cannot rely on the viewport centre only  to characterise the user behaviour. The  distance $r$ and the spatial coordinates on the virtual floor $x$ are also needed to formally define the navigation trajectory for a generic \sdof user $i$ as $ \{(x^i_1,p^i_1,r^i_1), (x^i_2,p^i_2,r^i_2), \dots,(x^i_n,p^i_n,r^i_n)\}$ \cite{rossi2021anewchallenge}. This information is crucial to define a simple similarity metric among users in this new setting.

\subsection{Overlap Ratio as the ground-truth metric}
\label{sec_overlapRatio}
Since we are interested in capturing viewers that are attending similar volumetric content at the same time instance, following the work presented in \cite{rossi2021anewchallenge}, the straightforward measure that could show this behaviour is the overlap among viewports. Given two users $i$ and $j$ shown in the right part of Figure~\ref{fig_6dofnavigation}, we denote their displayed viewport as $\mathcal{S}_t^i$ and $\mathcal{S}_t^j$, respectively, defined as the set of points of the volumetric content falling within their viewing frustum. Then, we denote the overlap set by $\mathcal{S}_t^{i} \cap \mathcal{S}_t^{j}$, the portion of points displayed by both users. Equipped with the above notation, we can now introduce a key metric for the analysis: the \textit{overlap ratio} $O(i,j)$. This is defined as the cardinality of the overlap set, normalised by the cardinality of the set containing all points of the volumetric content visualised by both users. More formally, the overlap ratio in a specific time $t$ is:
\begin{equation}
    O_t(i,j) = \frac{|\mathcal{S}_t^{i} \cap \mathcal{S}_t^{j}|}{|\mathcal{S}_t^{i} \cup \mathcal{S}_t^{i} |}
    \label{eq_overlapRatio}
\end{equation}
where $\mathcal{S}_t^i$ and $\mathcal{S}_t^j$ are the displayed viewport of users $i$ and $j$, respectively. In particular, a high value of overlap ratio means high similarity between users of the displayed content, and conversely. Even if this metric is exact and a clear indicator of how much similar users are with respect to their displayed content, its evaluation is not trivial as it is intensely time-consuming. For instance, the overlap ratio between two users requires 0.8986 
seconds per frame on average on an Intel R machine with CPU E5-4620 at 2.10 GHz; the operation needs to be computed for all the possible combinations of users, leading to a large overhead which does not meet requirements for real-time and scalable applications. A new measure is  needed to perform real-time applications. In the rest of the paper, we will use this metric as the ground truth of overlap among users and investigate different weights as a proxy for viewport overlap.
\subsection{Clustering as a tool for behavioural analysis}
\label{sec_clusteringAlgo}
Being able to assess users similarities in an objective way might be crucial for different applications such as behavioural analysis. As shown in \cite{rossi2019spherical}, a clique-based clustering algorithm is used to detect users with similar behaviour. This requires a reliable graph to be constructed in such a way that only the nodes that identify similar users (\ie who are displaying the same portion of the content) are connected. 
Equipped with such a meaningful graph, the clique-based clustering identifies optimal sub-graphs of all inter-connected nodes, ensuring the identification  of  the largest cluster of users all sharing a large viewport overlap.  In more detail, given a set of users who are experiencing the same content, we can represent their movements in a time-window $T$ as a set of graphs $\{ \mathcal{G}_t\}_{t=1}^T$. Each unweighted and undirected  graph $\mathcal{G}_t = \{\mathcal{V}, \mathcal{E}_t, \text{A}_t \}$ represents behavioural similarities among users at time $t$, where $\mathcal{V}$ and $\mathcal{E}_t$ denote the node and edge sets of $\mathcal{G}_t$, respectively. Each node in  $\mathcal{V}$ corresponds to a user interacting with the content. Each edge in $\mathcal{E}_t$ connects neighbouring nodes defined by the binary adjacency matrix $\text{A}_t$. Assuming that users are connected if they are displaying similar content, we can formally define the adjacency matrix $\text{A}_t$ as follow:
\begin{equation}
\label{eq:clustering_graph}
	\text{A}_t(i,j) = 	 
	\begin{cases} 
	1,    & \mbox{if } g_t(i,j) \geq G_{th} \\ 
	0, & \text{otherwise}. \\
	\end{cases} 
\end{equation}
where $g_t(i,j)$ is a similarity metric between user $i$ and $j$ and $G_{th}$ is a thresholding value. On this final graph, the clique-based clustering algorithm can be applied to identify a set of users all connected (\ie clique), and therefore with similar behaviour. In \cite{rossi2019spherical}, this  graph construction is based on a pairwise similarity metric specifically for the \tdof trajectories.  
\newline
Identifying a generic and reliable metric $g(i,j)$ that approximates behavioural similarities among users who experience a \sdof content is a key step to enable user behavioural analysis via tools proposed for \tdof scenario and the focus on the next section.

 \begin{table}[t]
 \caption{Definition of distance features and measurements.}
\label{table_definition_features}
\vspace{-0.8em} 
\centering
\resizebox{0.49\textwidth}{!}{
\begin{tabular}{@{\extracolsep{4pt}}ll}
\toprule   
{Symbol} & {Definition}\\
 \midrule
$x$ & user position on the VR floor \\ 
$p$ & viewport center projected on the volumetric content\\
$r$ & relative distance between user and volumetric content\\
$\mathsf{L}(\cdot,\cdot)$ & difference of relative distance between two users\\
$\mathsf{E}(\cdot,\cdot)$ & Euclidean distance\\ 
$\mathsf{G}(\cdot,\cdot)$ & Geodesic distance\\ 
\bottomrule
\end{tabular}} 
\vspace{-2em}
\end{table}

 \begin{table*}[t]
 \caption{Similarity metrics: definitions, included distance features and measurements, regulator and threshold values.}\label{table_metrics}
 \vspace{-0.8em}
\centering
\resizebox{1\textwidth}{!}{
\begin{tabular}{@{\extracolsep{4pt}}lllll}
\toprule   
{Symbol} & {Definition} & {Distance Feature and Metric} & {Regulator values} & {$S_{th}$} \\
 \midrule
$w_1$ & $k_{\alpha}^{(\mathsf{E})}(x^i,x^j)$ & $\mathsf{E}$($x^i,x^j$) & $\alpha$  = 1 & 0.64\\ 
$w_2$ & $k_{\alpha}^{(\mathsf{L})}(r^i,r^j) $ & $\mathsf{L}$($r^i,r^j$) & $\alpha$  = 1 & 0.80 \\  
$w_3$ & $k_{\alpha}^{(\mathsf{G})}(p^i,p^j)$& $\mathsf{G}$($p^i,p^j$) & $\alpha$  = 1 & 0.63 \\ 
$w_4$ & $k_{\alpha}^{(\mathsf{E})}(p^i,p^j)$ & $\mathsf{E}$($p^i,p^j$) & $\alpha$  = 1 & 0.84 \\ 
$w_5$ & $ k_{\alpha}^{(\mathsf{E})}(x^i,x^j) \cdot        k_{\beta}^{(\mathsf{L})}(r^i,r^j) \cdot k_{\gamma}^{(\mathsf{G})}(p^i,p^j)$ & $\mathsf{E}$($x^i,x^j$), $\mathsf{L}$($r^i,r^j$), $\mathsf{G}$($p^i,p^j$) & $\alpha$ = 0.1; \hspace{0.2em} $\beta$ = 0.5; \hspace{0.65em} $\gamma$ = 1 & 0.54\\ 
$w_6$ & $ k_{\alpha}^{(\mathsf{E})}(x^i,x^j) \cdot        k_{\beta}^{(\mathsf{L})}(r^i,r^j) \cdot k_{\gamma}^{(\mathsf{E})}(p^i,p^j)$ & $\mathsf{E}$($x^i,x^j$), $\mathsf{L}$($r^i,r^j$), $\mathsf{E}$($p^i,p^j$)  & $\alpha$ = 0.1; \hspace{0.2em} $\beta$ = 0.125; $\gamma$ = 0.2 & 0.87 \\ 
$w_7$ & $ k_{\alpha}^{(\mathsf{E})}(x^i,x^j) \cdot       \beta [\eta(r_i)+\eta(r_j)] \cdot
        k_{\gamma}^{(\mathsf{G})}(p^i,p^j)$ & $\mathsf{E}$($x^i,x^j$), $r^i$,$r^j$, $\mathsf{G}$($p^i,p^j$)  & $\alpha$ = 0.25;  $\beta$ = 0.5; \hspace{0.65em} $\gamma$ = 0.5  & 0.60\\ 
$w_8$ & $k_{\alpha}^{(\mathsf{E})}(x^i,x^j) \cdot       \beta [\eta(r_i)+\eta(r_j)] \cdot
        k_{\gamma}^{(\mathsf{E})}(p^i,p^j)$  & $\mathsf{E}$($x^i,x^j$), $r^i,r^j$, $\mathsf{E}$($p^i,p^j$)   &$\alpha$ = 0.5; \hspace{0.2em} $\beta$ = 0.5;  \hspace{0.65em} $\gamma$ = 0.5 & 0.62\\ 
\bottomrule
\end{tabular}} 
\vspace{-1.2em}
\end{table*}

\section{Proposed metrics}
\label{sec_proposed_metrics}
In this section, we present eight similarity metrics and we provide an exhaustive study to understand which one approximates at the best the viewport overlap. Those metrics are expressed as a function of  
various  \textit{distance features} and \textit{measurements} 
considering either users' position on the floor ($x$) or users' viewing direction in terms of the viewport centre projected on the volumetric content ($p$) or both. We divide the metrics into two groups: \emph{single-feature}   and \emph{multi-feature} metrics. For the sake of notation, we omit the temporal parameter $t$. Table~\ref{table_definition_features} summarises the distance features and measurements that we consider, while our proposed similarity metrics are reported in Table~\ref{table_metrics}. 

\subsection{Single-feature metrics to assess users similarity}
The first set of similarity metrics is based on one single distance feature. 
We model the similarity functions via radial basis function kernel. Specifically, we consider the Gaussian kernel \cite{stankovic2019understanding} defined  as follows:
\begin{equation} 
    k_{\alpha}^{(D)}(i,j) =  e^{-{\alpha D(i,j)}}
\end{equation}
where  $D(i,j)$ is the distance between two generic users $i$ and $j$, while $\alpha > 0$ is a parameter to better regularise the distance. This distance can be evaluated in multiple ways and we consider the distance features and measurements taken into account in \cite{rossi2021anewchallenge}. Specifically, the first two similarity metrics $w_1$ and $w_2$ are based on the location of users in the virtual space with respect to the virtual object or other viewers. The former metric is based on the Euclidean distance $\mathsf{E}(x^i,x^j)$ between user $i$ and $j$ on the virtual floor. Instead, $w_2$  considers the difference in terms of the relative distance of users to the centroid of the displayed content, $\mathsf{L} = ||r^i - r^j||$. Specifically, we define them as follows:
\begin{align} 
    w_1 &=   e^{-\alpha \mathsf{E}(x^i,x^j)} = k_{\alpha}^{(\mathsf{E})}(x^i,x^j);  \\
    w_2 &=  e^{-\alpha ||r^i - r^j||} = k_{\alpha}^{(\mathsf{L})}(r^i,r^j).
\end{align}
\\
The metrics $w_3$ and $w_4$ are instead based on the distance between the two viewport centres $p$ of user $i$ and  user $j$ projected on the volumetric content. To take into account the heterogeneous shape of the volumetric content, this distance in $w_3$ is measured in terms of the Geodesic distance $\mathsf{G}(p^i,p^j)$ while in $w_4$ in terms of the Euclidean distance $\mathsf{E}(p^i,p^j)$. More formally, they are defined as:
\begin{align} 
    w_3 &= k_{\alpha}^{(\mathsf{G})}(p^i,p^j) =  e^{-\alpha \mathsf{G}(p^i,p^j)}
    \\
    w_4 &= k_{\alpha}^{(\mathsf{E})}(p^i,p^j) =  e^{-\alpha \mathsf{E}(p^i,p^j)}.
\end{align}

\subsection{Multi-feature metrics to assess users similarity}
As emerged in \cite{rossi2021anewchallenge}, both user viewing direction and position on the virtual floor are relevant to detect similar behaviour among users. Thus, the last set of proposed similarity metrics considers a combination of distance features. In detail, $w_5$ and $w_6$ are based on the previous similarity metrics $w_1$ and $w_2$, but include also the distance of their viewport centres $p$ projected on the volumetric content in terms of Geodesic distance $\mathsf{G}(p^i,p^j)$ and Euclidean distance $\mathsf{E}(p^i,p^j)$, respectively. More formally, we define $w_5$ as:
\begin{equation}
\begin{split}
    w_5 &= k_{\alpha}^{(\mathsf{E})}(x^i,x^j) \cdot        k_{\beta}^{(\mathsf{L})}(r^i,r^j) \cdot
        k_{\gamma}^{(\mathsf{G})}(p^i,p^j) \\
        &= e^{-\alpha \mathsf{E}(x^i,x^j)}\cdot e^{-\beta ||r^i - r^j||}\cdot e^{-\gamma \mathsf{G}(p^i,p^j)}; \\
\end{split}
\end{equation} 
while the second weight is equal to:
\begin{equation}
\begin{split}
    w_6 &= k_{\alpha}^{(\mathsf{E})}(x^i,x^j) \cdot        k_{\beta}^{(\mathsf{L})}(r^i,r^j) \cdot
        k_{\gamma}^{(\mathsf{E})}(p^i,p^j) \\
        &= e^{-\alpha \mathsf{E}(x^i,x^j)}\cdot e^{-\beta ||r^i - r^j||}\cdot e^{-\gamma \mathsf{E}(p^i,p^j)}. \\
\end{split}
\end{equation} 
For the sake of clarity, $\beta$ and $\gamma$ are regulators such as $\alpha$.\\
The preliminary analysis presented in \cite{rossi2021anewchallenge} has also highlighted a correlation between the viewport overlap of two users and their relative distance from the volumetric content.  The closer users are to the volumetric content, the smaller and more detailed is the portion of the displayed content; the farther they are, the bigger but with fewer details becomes the displayed portion. Thus, in the first case, the high overlap between displayed areas of two different users is more difficult. To take into consideration this behaviour, we model the relative distance via a hyperbolic tangent kernel. Given the relative distance $r_i$ between the user $i$ and volumetric content, we evaluate it as follows:
\begin{equation}
    \eta(r_i) = \tanh\left(r_i\right).
\end{equation}
As previously, metrics $w_7$ and $w_8$ are based on both user distance in the virtual floor $\mathsf{E}(x^i,x^j)$, and on the volumetric content in terms of Geodesic distance $\mathsf{G}(p^i,p^j)$ and Euclidean distance $\mathsf{E}(p^i,p^j)$, respectively. 
More formally, we define $w_7$ as follows:
\begin{equation}
\begin{split}
    w_7 &= k_{\alpha}^{(\mathsf{E})}(x^i,x^j) \cdot \beta \big[ \eta(r^i) + \eta(r^j) \big] \cdot
        k_{\gamma}^{(\mathsf{G})}(p^i,p^j) \\
        &= e^{-\alpha \mathsf{E}(x^i,x^j)}\cdot \beta \big[ \tanh\left(r_i\right) + \tanh\left(r_j\right) \big] \cdot e^{-\gamma \mathsf{G}(p^i,p^j)}; 
\end{split}
\end{equation} 
while $w_8$ is:
\begin{equation}
\begin{split}
    w_8 &= k_{\alpha}^{(\mathsf{E})}(x^i,x^j) \cdot \beta \big[ \eta(r^i) + \eta(r^j) \big] \cdot
        k_{\gamma}^{(\mathsf{E})}(p^i,p^j) \\
        &= e^{-\alpha \mathsf{E}(x^i,x^j)}\cdot \beta \big[ \tanh\left(r_i\right) + \tanh\left(r_j\right) \big] \cdot e^{-\gamma \mathsf{E}(p^i,p^j)}. 
\end{split}
\end{equation}

\section{Experimental setup}
\label{sec_validation}
We now validate the above metrics using a point cloud dataset. 
We now describe the navigation dataset and how we evaluate the performance of our similarity metrics (Section~\ref{sec_dataset} and \ref{sec_performance}, respectively). Then, we run an ablation study to evaluate for each similarity metric the best-performing set of regulators. 

\subsection{Dataset and Methodology}
\label{sec_dataset}
\begin{figure}[t]
    \centering
	\subfigure[\textit{Long Dress}  (PC 1)]{\includegraphics[height=0.25\textwidth]{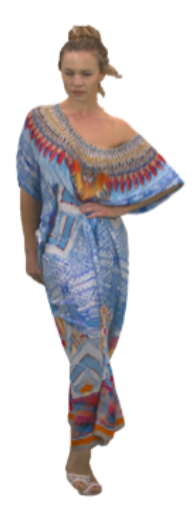}}
    \quad
    \subfigure[\textit{Loot} \hspace{1em} (PC 2)]{\includegraphics[height=0.25\textwidth]{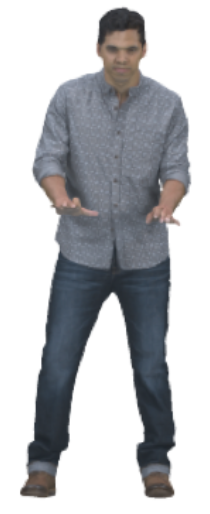}}
    \quad
    \subfigure[\textit{Red and Black} (PC 3)]{\includegraphics[height=0.25\textwidth]{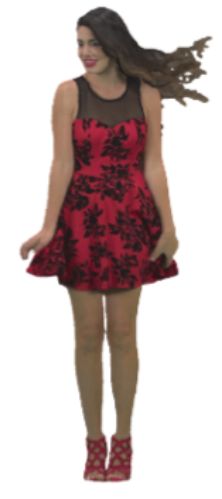}}
    \quad
    \subfigure[\textit{Soldier} (PC 4)]{\includegraphics[height=0.25\textwidth]{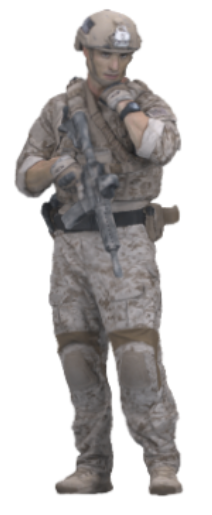}}
    \vspace{-1.5em}
    \caption{Human Body Point Clouds \cite{krivokuca20188i} content used in the collection of a public available dataset presented in \cite{subramanyam2020user}.}
    \vspace{-1em}
    \label{Fig_HumanBodyDataset}
\end{figure}
{\bf Dataset.}
Existing datasets with user navigation collected  while displaying volumetric objects in a \sdof environment are still very limited. In the following, we use  the open dataset presented in~\cite{subramanyam2020user}. 
This is comprised of navigation trajectories of 26 users participating in a visual quality assessment study in VR. For the study, four dynamic point cloud sequences were employed~\cite{krivokuca20188i}, namely \textit{Long dress} (PC1), \textit{Loot} (PC2), \textit{Red and black} (PC3), \textit{Soldier} (PC4) (Figure~\ref{Fig_HumanBodyDataset}). Each sequence was distorted at four different bit rate points with two compression algorithms: the anchor used for the MPEG call for proposals, and the upcoming MPEG standard V-PCC. Hidden references were additionally employed in the test, for a total of 36 stimuli. Similarly to what is shown in Figure~\ref{fig_6dofnavigation}, a single object of interest was placed in the \ac{VR} scene, and users were instructed to focus on the volumetric content for the duration of the session and rate its visual quality. Therefore, the navigation data adheres to the assumptions listed in Section~\ref{sec_challenges_new}.
\newline
{\bf Graph Construction.}
To implement the graph-based clustering proposed in \cite{rossi2019spherical} based on our proposed similarity metrics, we need to construct a binary graph following Equation~\eqref{eq:clustering_graph}, as  described in Section~\ref{sec_clusteringAlgo}. To be noted, our proposed similarity metrics are based on distance measurements. As shown in \cite{rossi2019spherical}, the correlation between overlap and distance is inversely proportional. This means that high values of overlap (and thus, high similarity) correspond to low distance. Therefore, the condition to construct the adjacency metric $\text{A}_t$ based on our proposed similarity metrics becomes the following: $w(i,j) \leq S_{th}$ where $w(i,j)$ is one of the similarity metrics proposed in Section~\ref{sec_proposed_metrics} and $S_{th}$ a threshold value which identifies similar users and thus, neighbours on the graph.
In short, users with a similarity metric below a threshold value $S_{th}$ are neighbours in the graph. Hence, the first step now is to identify $S_{th}$. Per each proposed similarity metric, we empirically evaluate the \ac{ROC}  curves based on the navigation trajectories of the entire dataset above described and select the best value of threshold as originally done in \cite{rossi2019spherical}.  Specifically, we set the thresholding values such that a good trade-off between True Positive Rate (TPR) and False Positive Rate (FPR) is met. As ground truth for the ROC,  we assumed that two users are attending the same portion of the content, and thus are classified as similar, if their viewports overlap by at least 75\% of their total viewed area. The predicted event is instead evaluated using the eight metrics presented in the previous section, and the corresponding threshold values are selected in order to have TPR  equal to $0.75$. For the sake of clarity, the ground-truth value of viewport overlap has been set equal to 75\% because ensures per each similarity metric a low probability to have a wrong classification (\ie FPR below $0.4$) without compromising the probability of correctly classifying the similarity event (\ie TPR) which remains above $0.75$. In the last column of Table~\ref{table_metrics}, we provide the selected $S_{th}$ per each similarity metric that will be used in the following. 
 \begin{table*}[ht!]
\centering
\vspace{-1em}
\caption{Parameter selections and their performance for multi-feature metrics ($w_5$ - $w_8$).}\label{table_metrics_multi_feature}
\vspace{-1em}
\resizebox{0.8\textwidth}{!}{
\begin{tabular}{@{\extracolsep{1pt}} c | c | cccc |}
 \multicolumn{1}{l}{} & \multicolumn{1}{l}{} & {$w_5$} & {$w_6$} & {$w_7$} & \multicolumn{1}{c}{ {$w_8$}} \\ 
\cline{2-6}
\multicolumn{1}{c}{\multirow{5}{*}{\rotatebox[origin=c]{90}{\textbf{\textcolor{set1}{set 1}}}}} & \multicolumn{1}{| c |}{ [$\alpha, \beta, \gamma $] } & [0.12, 0.125, 0.125] & [0.12, 1, 0.25] & [0.125, 0.5, 0.25] & [0.25, 0.5, 0.2]  \\
\multicolumn{1}{ c}{} & \multicolumn{1}{| c |}{Overlap Ratio} & 0.63 & 0.64 & 0.66 & \textbf{0.69} \\
\multicolumn{1}{ c}{} & \multicolumn{1}{| c |}{ Relevant Population} & 0.82 & 0.78 & 0.69 & 0.62 \\
\multicolumn{1}{ c}{} & \multicolumn{1}{| c |}{Precision} & 0.45 & 0.40 & 0.47 & 0.48  \\
\cline{2-6}
\multicolumn{1}{c}{\multirow{5}{*}{\rotatebox[origin=c]{90}{\textbf{\textcolor{set2}{set 2}}}}} & \multicolumn{1}{| c |}{ [$\alpha, \beta, \gamma $] } &  [1, 0.05, 0.05] & [0.5, 0.05, 0.05] & [2, 0.5, 0.1] & [2, 0.5, 0.05]  \\
\multicolumn{1}{c}{} & \multicolumn{1}{| c |}{Overlap Ratio} &  0.58 & 0.59 & 0.60 & 0.63   \\
\multicolumn{1}{ c}{} & \multicolumn{1}{| c |}{ Relevant Population}  & \textbf{0.91} & 0.89 & 0.87 & 0.84  \\
\multicolumn{1}{ c}{} & \multicolumn{1}{| c |}{Precision} & 0.32 & 0.32 & 0.36 & 0.33     \\
\cline{2-6}
\multicolumn{1}{c}{\multirow{5}{*}{\rotatebox[origin=c]{90}{\textbf{\textcolor{set3}{set 3}}}}} & \multicolumn{1}{| c |}{ [$\alpha, \beta, \gamma $] }   & [0.1, 0.5, 1] & [0.1, 0.125, 0.2] & [0.25, 0.5, 0.5] & [0.5, 0.5, 0.5]  \\
\multicolumn{1}{ c}{} & \multicolumn{1}{| c |}{Overlap Ratio} & 0.63 & 0.63 & 0.65 & 0.66  \\
\multicolumn{1}{ c}{} & \multicolumn{1}{| c |}{Relevant Population}  & 0.83 & 0.80 & 0.77 & 0.74  \\
\multicolumn{1}{ c}{} & \multicolumn{1}{| c |}{Precision}  & 0.45 & 0.44 & \textbf{0.49} & 0.48    \\
\cline{2-6}
\end{tabular}}
\vspace{-1em}
\end{table*}

\begin{figure*}[t]
	\centering
	\subfigure[Overlap Ratio]{
	\includegraphics[width=0.32\textwidth]{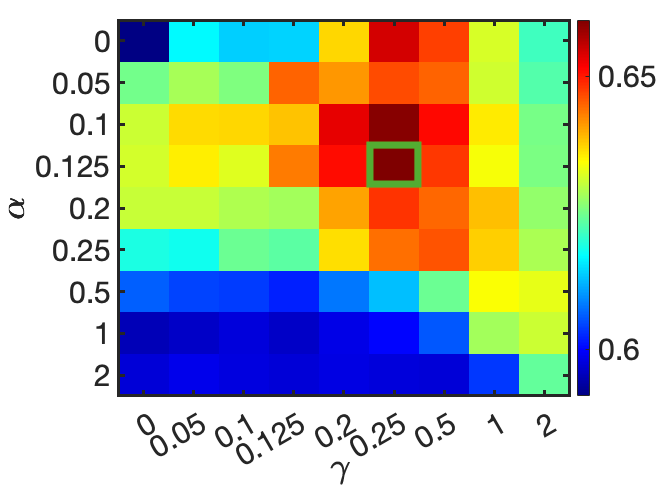}}
	\quad
    \hspace{-1em}
	\subfigure[Relevant population]{
		\includegraphics[width=0.32\textwidth]{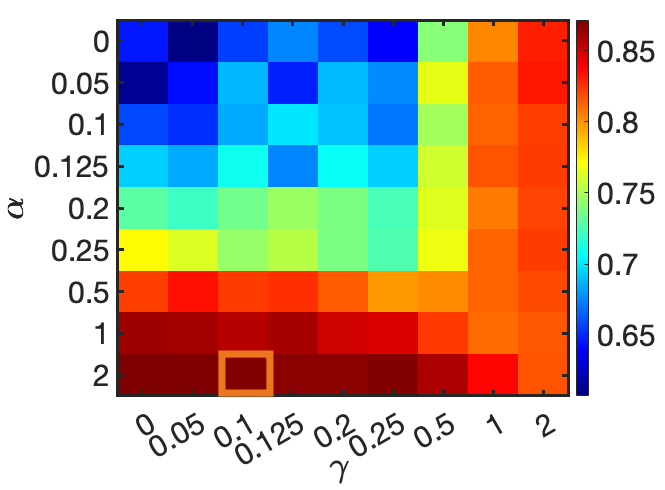}}
	\quad
    \hspace{-1em}
	\subfigure[Precision]{ 
		\includegraphics[width=0.32\textwidth]{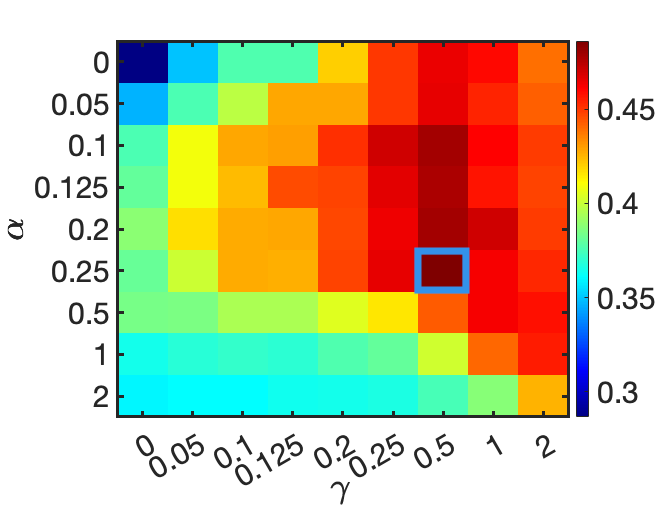}}
  \vspace{-1em}
	\caption{Example of parameter selection for $w_7$ with $\beta = 0.5$. Values \textcolor{set1}{set 1} selected based on max overlap, \textcolor{set2}{set 2} max clustered users, \textcolor{set3}{set 3} based on precision.}
	\label{Fig_parameter_selection_w8}
 \vspace{-1em}
\end{figure*}
\subsection{Performance Evaluation Setup}
\label{sec_performance}
To validate our proposed similarity  metrics, we consider three performance metrics: averaged \emph{overlap ratio} per cluster, \emph{relevant clustered population}, and \emph{precision}. The first two are more specific to our navigation trajectory in a \ac{VR} system, while the latter is a popular index used to evaluate clustering algorithm performance. \newline 
\textbf{Overlap ratio per cluster}: as defined in Section~\ref{sec_overlapRatio}, the overlap ratio computes the portion in common of displayed content between two users. Therefore, to compare the performance of our detected clusters with the different similarity metrics, we average the overlap ratio among all users who are put in the same group. More formally, given a detected cluster $C_k$ is defined as follows:
\begin{equation}
O_k = \frac{\displaystyle 1}{n_k}   \sum_{\substack{i,j\in C_k\\ i\neq j}} O(i,j) 
\vspace{-0.8em}
\end{equation}
where $i$ and $j$ are two generic users, $n_k$ is the cardinality of elements bellowing to cluster $C_k$ and $O(i,j)$ the overlap ratio as in Equation~\ref{eq_overlapRatio}.  \newline
\textbf{Relevant clustered population}: the more users are clustered together with high viewport overlap, the more meaningful are our clusters. Thus, we consider as relevant clustered population the sum of users that have been put in clusters with more than 2 elements. \newline
\textbf{Precision}: in a classification task, this index evaluates the portion of elements that are classified correctly and has values between 0 and 1 \cite{fawcett2006introduction}. More formally:
\begin{equation}
    P = \frac{TP}{TP+FP}
\end{equation}
where \ac{TP} (\ac{FP})  is the number of viewers classified correctly (incorrectly) together in a cluster.  In our case, two users are identified positively if they are in the same cluster and their viewport overlap is actually over the desired threshold.
\begin{figure*}[ht!]
	\centering
    \vspace{-.2cm}
	\hspace{-2em}
	\subfigure[Ground-truth ($O_{th}$ = 75\%)]{
		\includegraphics[width=0.34\textwidth]{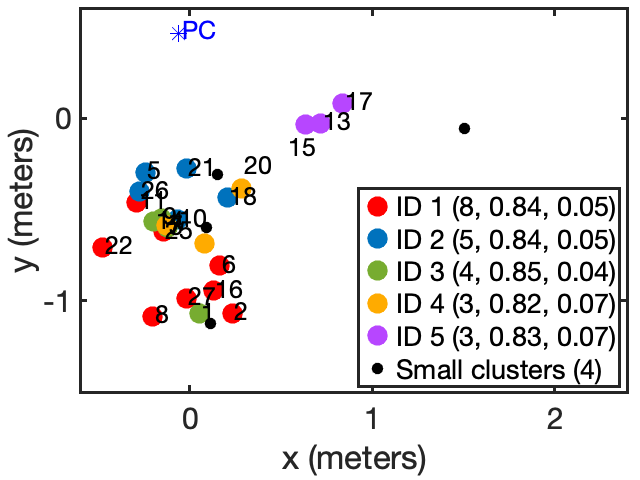}}
	\quad
    \hspace{-2em}
	\subfigure[$w_1$ (single feature metric)]{ 
		\includegraphics[width=0.34\textwidth]{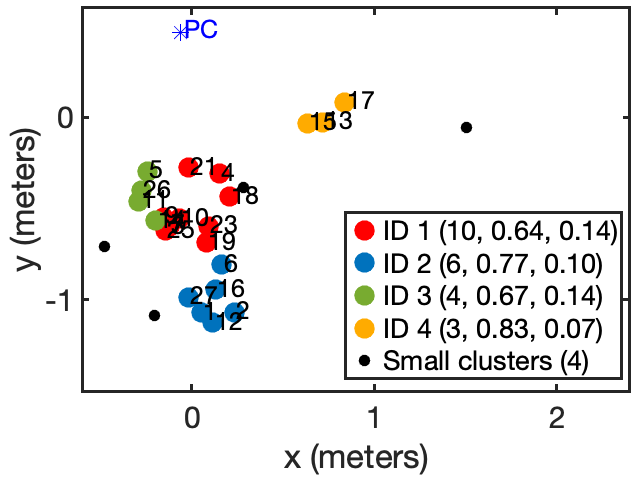}}
    \quad
    \hspace{-2em}
	\subfigure[$w_2$ (single feature metric)]{
		\includegraphics[width=0.34\textwidth]{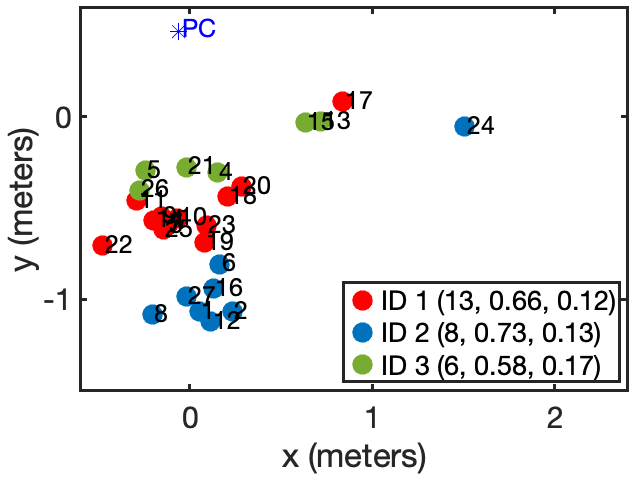}}
	\\
    \vspace{-.2cm}
    \hspace{-2em}
	\subfigure[$w_3$ (single feature metric)]{
		\includegraphics[width=0.34\textwidth]{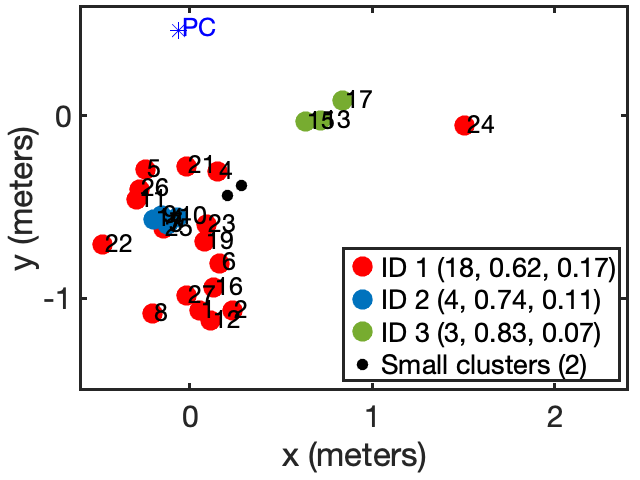}}
	\quad
    \hspace{-2em}
	\subfigure[$w_4$ (single feature metric)]{
		\includegraphics[width=0.34\textwidth]{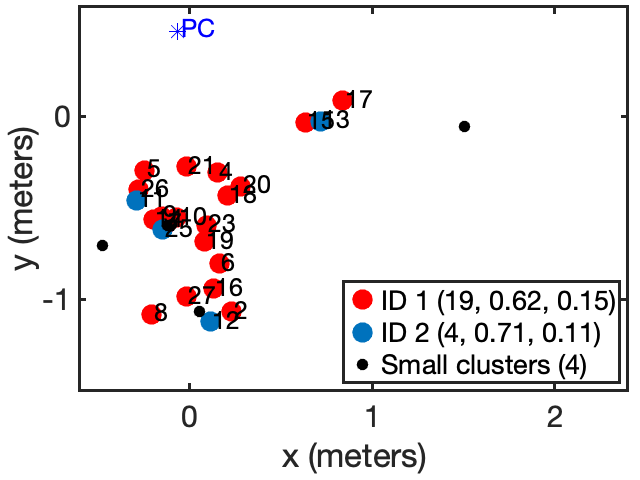}}
		 \hspace{-0.9em}
	\subfigure[$w_5$ (multi-feature metric)]{
		\includegraphics[width=0.34\textwidth]{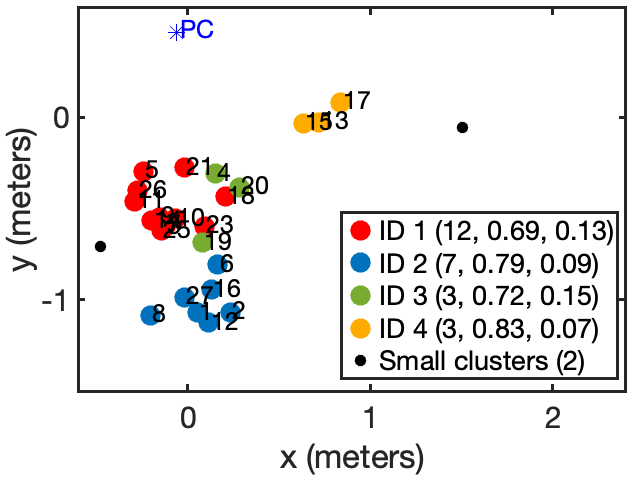}}
	\\
    \vspace{-.2cm}
    \hspace{-2.2em}
	\subfigure[$w_6$ (multi-feature metric)]{ 
		\includegraphics[width=0.34\textwidth]{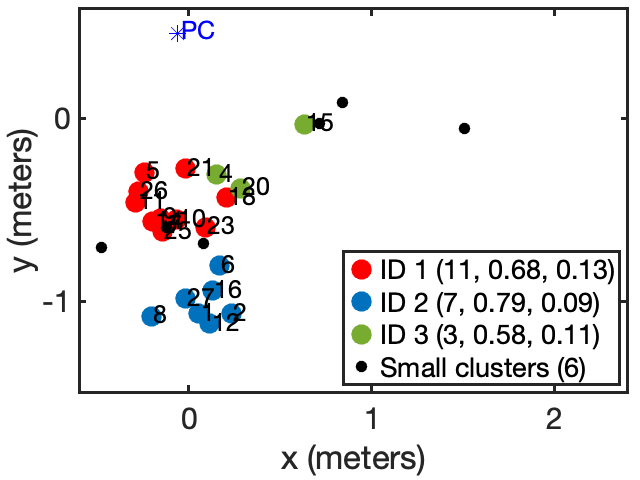}}
    \quad
    \hspace{-2em}
	\subfigure[$w_7$ (multi-feature metric)]{
		\includegraphics[width=0.34\textwidth]{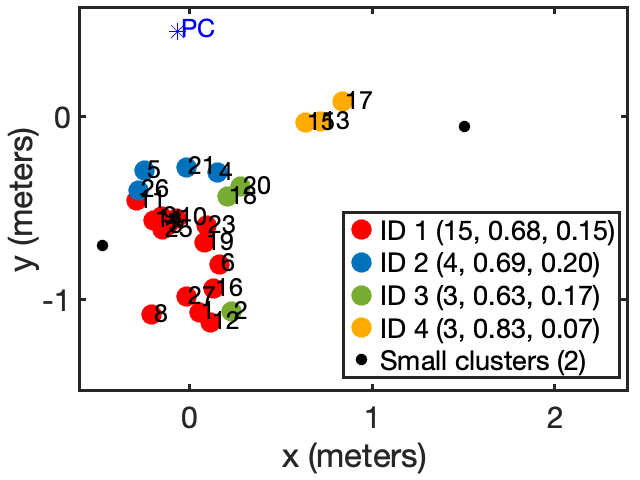}}
	\quad
    \hspace{-1.8em}
	\subfigure[$w_8$ (multi-feature metric)]{
		\includegraphics[width=0.34\textwidth]{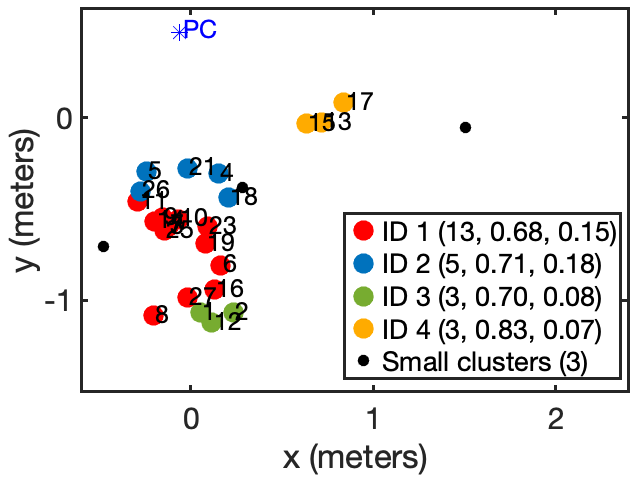}}
		\vspace{-1em}
	\caption{Cluster results in frame 50 of sequence PC1 (\textit{Longdress}). Each dot represents a user on the virtual floor while the blue star stands for the volumetric content. In the legend in brackets, per each cluster with more than 2 users are reported: the number of users in the same cluster, averaged pairwise viewport overlap and corresponding variance within the cluster.}
	\vspace{-1em}
	\label{Fig_clusteringFloor}
\end{figure*}
\subsection{Ablation Study}
\label{sec_ablationStudy}
We finally present an ablation study to tune the best set of regulator parameters that maximise the performance of each similarity metric. Equipped with the threshold values given in Table~\ref{table_metrics}, we run a frame-based clustering to select the best regulators $\alpha$, $\beta$ and $\sigma$. We test their performance in terms of the metrics above described in the following range of values $[0, 0.05, 0.1, 0.125, 0.2, 0.25, 0.5, 1, 2]$ based on navigation trajectories collected in the entire dataset. Finally, we average over time and across content the performance of each cluster obtained by all the similarity metrics.\\
\textbf{Single-feature metrics}\\
For single-feature metrics ($w_1-w_4$), we notice a very small variance in terms of performance. Thus, we selected $\alpha = 1$ for this set. \\
\textbf{Multi-feature metrics}\\
More challenging is instead the selection parameters for multi-feature metrics ($w_5-w_8$). Each similarity metric depends on three parameters: $\alpha$, $\beta$ and $\gamma$. To overcome this, we first select three sets of parameters taking into account only navigation trajectories for reference content: one group of parameters (\textcolor{set1}{set 1}) based on the maximum overlap ratio, the second (\textcolor{set2}{set 2}) on the maximum relevant clustered population and the last group (\textcolor{set3}{set 3}) as the one reaching the highest precision. As an example, Figure~\ref{Fig_parameter_selection_w8} shows the selection of these three sets of parameters for the metric $w_7$. Then, we test these on all the available trajectories included in the analysed dataset to finally select the best set of parameters. Table~\ref{table_metrics_multi_feature} provides all the performance of the multi-feature similarity metrics obtained by the three selected sets of parameters. Since there is no particular configuration that outperforms in terms of overlap ratio, relevant population and precision, we decided to select \textcolor{set3}{set 3}. This configuration, besides ensuring the highest value of precision, also guarantees acceptable values of overlap ratio and relevant population for all the similarity metrics. For example for $w_7$, selecting values of \textcolor{set3}{set 3} means that users are correctly clustered in almost the 50$\%$ of the time (precision equal to 0.49); at the same time the 77$\%$ of the population is put in clusters with more than the 2 users (relevant population equal to 0.77) and on average the overlap of viewport between users in the same cluster is consistent (overlap ration equal to 65$\%$). These values are similar to the highest value for $w_7$ of the relevant population and overlap ratio which are $0.87$ and $0.66$, respectively. Table~\ref{table_metrics} summarises the values used in the following.
\section{Results}
\label{sec_analysis_results}
\begin{table*}[t]
\caption{Results in terms of averaged and standard deviation per each performance metric across the entire dataset.}\label{table_performance_metrics_2}
\vspace{-0.8em}
\centering
\resizebox{\textwidth}{!}{
\begin{tabular}{ c|c| cccc cccc}
\toprule
\multicolumn{1}{ c }{ } & \multicolumn{1}{ c }{\textbf{Metrics}} & $w_1$ & $w_2$ & $w_3$ & $w_4$ & $w_5$ & $w_6$ & $w_7$ & $w_8$ \\

\midrule
\multirow{3}{*}{
     \centering \rotatebox{90}{\textbf{PC1}}} & Overlap Ratio & 0.68 $\pm$ 0.05 & 0.65 $\pm$ 0.04 & 0.66 $\pm$ 0.04 & 0.68 $\pm$ 0.07 & 0.70 $\pm$ 0.05 & 0.71 $\pm$ 0.05 & 0.70 $\pm$ 0.05  & \textbf{0.72 $\pm$ 0.06} \\
     & Relevant Population & 0.85 $\pm$ 0.04 & \textbf{0.94 $\pm$ 0.03} & 0.92 $\pm$ 0.05 & 0.84 $\pm$ 0.08 & 0.83 $\pm$ 0.06 & 0.83 $\pm$ 0.07 & 0.83 $\pm$ 0.06 & 0.83 $\pm$ 0.07 \\
    & Precision & 0.44 $\pm$ 0.06 & 0.35 $\pm$ 0.05 & 0.39 $\pm$ 0.07 & 0.30 $\pm$ 0.06 & 0.47 $\pm$ 0.07 & \textbf{0.49 $\pm$ 0.08} & 0.46 $\pm$ 0.07 & 0.44 $\pm$ 0.10 \\
\hline
\multirow{3}{*}{
     \centering \rotatebox{90}{\textbf{PC2}}} & Overlap Ratio  & 0.57 $\pm$ 0.08 & 0.53 $\pm$ 0.09 & 0.54 $\pm$ 0.12 & 0.54 $\pm$ 0.11 & 0.59 $\pm$ 0.08 & 0.58 $\pm$ 0.08 & 0.59 $\pm$ 0.12 & \textbf{0.60 $\pm$ 0.10}\\ 
     & Relevant Population & 0.80 $\pm$ 0.07 & \textbf{0.92 $\pm$ 0.06} & 0.83 $\pm$ 0.07 & 0.89 $\pm$ 0.06 & 0.80 $\pm$ 0.10 & 0.81 $\pm$ 0.07 & 0.72 $\pm$ 0.08 & 0.73 $\pm$ 0.06 \\
    & Precision & 0.45 $\pm$ 0.06 & 0.28 $\pm$ 0.08 & 0.31 $\pm$ 0.08 & 0.27 $\pm$ 0.08 & 0.47 $\pm$ 0.09 & 0.42 $\pm$ 0.08 & \textbf{0.54 $\pm$ 0.08} & \textbf{0.54 $\pm$ 0.12} \\
\hline    
\multirow{3}{*}{
     \centering \rotatebox{90}{\textbf{PC3}}} & Overlap Ratio & 0.65 $\pm$ 0.06 & 0.60 $\pm$ 0.07 & 0.64 $\pm$ 0.05 & 0.68 $\pm$ 0.06 & 0.65 $\pm$ 0.06 & 0.65 $\pm$ 0.06 & 0.68 $\pm$ 0.05 & \textbf{0.69 $\pm$ 0.05} \\
     & Relevant Population & 0.82 $\pm$ 0.07 & \textbf{0.93 $\pm$ 0.05} & 0.88 $\pm$ 0.06 & 0.82 $\pm$ 0.08 & 0.84 $\pm$ 0.06 & 0.81 $\pm$ 0.07 & 0.72 $\pm$ 0.07 & 0.70 $\pm$ 0.07 \\
    & Precision & 0.48 $\pm$ 0.11 & 0.36 $\pm$ 0.08 & 0.39 $\pm$ 0.07 & 0.39 $\pm$ 0.06 & 0.49 $\pm$ 0.11 & 0.49 $\pm$ 0.10 & 0.52 $\pm$ 0.08 & \textbf{0.55 $\pm$ 0.08} \\
\hline  
\multirow{3}{*}{
     \centering \rotatebox{90}{\textbf{PC4}}}  & Overlap Ratio & 0.60 $\pm$ 0.04 & 0.52 $\pm$ 0.06 & 0.55 $\pm$ 0.03 & 0.59 $\pm$ 0.06 & 0.59 $\pm$ 0.04 & 0.58 $\pm$ 0.05 & 0.61 $\pm$ 0.04 & \textbf{0.66 $\pm$ 0.05} \\
     & Relevant Population & 0.82 $\pm$ 0.07 & \textbf{0.92 $\pm$ 0.05} & 0.90 $\pm$ 0.08 & 0.86 $\pm$ 0.08 & 0.83 $\pm$ 0.08 & 0.77 $\pm$ 0.07 & 0.80 $\pm$ 0.07 & 0.71 $\pm$ 0.08 \\
    & Precision & 0.35 $\pm$ 0.06 & 0.22 $\pm$ 0.04 & 0.31 $\pm$ 0.06 & 0.25 $\pm$ 0.07 & 0.38 $\pm$ 0.07 & 0.38 $\pm$ 0.09 & \textbf{0.42 $\pm$ 0.06} & \textbf{0.42 $\pm$ 0.07} \\
\hline  
\hline
\multirow{3}{*}{
     \rotatebox{90}{\textbf{ All PCs\hspace{0.01cm}}}} & Overlap Ratio & 0.62 $\pm$ 0.06  & 0.57 $\pm$ 0.06  & 0.60 $\pm$ 0.06  & 0.62 $\pm$ 0.07  & 0.63 $\pm$ 0.06  & 0.63 $\pm$ 0.06  & 0.65 $\pm$ 0.06  & \textbf{0.66 $\pm$ 0.06}\\
     & Relevant Population & 0.82 $\pm$ 0.06  &  \textbf{0.93 $\pm$ 0.05} & 0.88 $\pm$ 0.07 & 0.85 $\pm$ 0.08 & 0.83 $\pm$ 0.07 & 0.80 $\pm$ 0.07 & 0.77 $\pm$ 0.07 & 0.74 $\pm$ 0.07 \\
    & Precision & 0.43 $\pm$ 0.07 & 0.30 $\pm$ 0.06 & 0.35 $\pm$ 0.07 & 0.30 $\pm$ 0.07 & 0.45 $\pm$ 0.09 & 0.45 $\pm$ 0.09 & \textbf{0.49 $\pm$ 0.07} & 0.48 $\pm$ 0.09 \\
\bottomrule  
\end{tabular}
}
\vspace{-1em}
\end{table*}

Equipped with the similarity metrics, the corresponding values of regulators and thresholds in Table~\ref{table_metrics}, we now conduct our validation study, focusing on analysing navigation trajectories experienced with non-distorted content. 
\subsection{Frame-Based Analysis}
As first step, we implement a frame-based analysis (i.e., frame-based clustering) to visually compare the detected clusters by the different similarity metrics. Figure~\ref{Fig_clusteringFloor} shows the clusters detected using the ground-truth metric $O$ to construct the graph (Figure~\ref{Fig_clusteringFloor}~(a)) with the ones given based on each proposed similarity metric (Figure~\ref{Fig_clusteringFloor}~(b-i)), for frame 50 of sequence PC1. In particular, each user is represented by a point on the VR floor which is coloured based on the assigned ID cluster, whereas the volumetric content is symbolised by a blue star. Per each relevant cluster (\ie cluster with more than 2 users), we provide in the legend the following results: the number of users inside the cluster, the average and variance of the overlap ratio among all users within the cluster. Finally, we represent the remaining users which are in either single or couple-cluster as black points; the total number of these users is also provided in the legend as ``Small clusters (total number of non-relevant clusters)".

Figure~\ref{Fig_clusteringFloor}~(a) shows the clusters that we consider as our ground truth since they are evaluated considering the overlap ratio $O$ as a similarity metric. In this case, 5 main clusters are detected with an average overlap ratio per cluster above 0.82. In particular, cluster ID 1 has the highest number of users (8) but has a relevant value of overlap ratio (0.84). Only 4 users in this case are put in single clusters. The goal is to find a similarity metric that can detect similar results. We can notice that single feature metrics, Figure~\ref{Fig_clusteringFloor}~(b-e), have the tendency to create very populated clusters but with a low overlap ratio. For instance, $w_3$ and $w_4$ generate a main big cluster with $18$ and $19$ users, respectively, while the corresponding overlap ratio drops drastically to $0.62$. The only exception is given by $w_1$, which generates a variable set of clusters with consistent values of overlap ratio, over $0.64$.
Let us now consider as an example the users $13, 15$ and $17$, which in the ground-truth case (Figure~\ref{Fig_clusteringFloor}~(a))  form their own cluster (\ie ID $5$)  with a high overlap ratio ($0.83$), and user $24$, who is quite isolated from other users and belongs to a single cluster. We can notice that $w_2$ and $w_4$ fail in detecting the group of users $13$, $15$ and $17$ as similar, dividing them instead in different clusters. On the other hand, $w_3$ detects this similarity but puts user $24$ in a relevant cluster (ID $1$). 
From these observations, we can notice that the viewport centre on the volumetric content, on which $w_3$ and $w_4$ are based, is not sufficient to correctly identify similar users. Analogously, considering only the difference in terms of the relative distance between the user and volumetric content, as done in $w_2$, does not allow the detection of similarity among users. Thus, the most promising metric in this group seems to be $w_1$, which is based on the user position on the virtual floor. 
\\
The last group of Figure~\ref{Fig_clusteringFloor}~(f-i) shows clusters based on multi-feature similarity metrics. In all these settings, a total of four main clusters are detected, except for $w_6$ which leads to three clusters, as shown in  Figure~\ref{Fig_clusteringFloor}~(g). The latter detects the highest number of small clusters ($6$) while being the only one that does not identify users $13$, $15$ and $17$ within the same cluster. 
On the contrary, the other three metrics $w_5$, $w_7$ and $w_8$ detect a main cluster and three smaller clusters with a consistent overlap ratio. For instance, the resulting clusters based on $w_5$ have an overlap ratio always bigger than 0.69 and only two users fall into small clusters. Overall, multi-functional metrics appear to be better suited to detect similar users than previous ones, with the exception of $ w_6 $. 

In Table~\ref{table_performance_metrics_2}, we extend our per-frame clustering analysis to the entire dataset: we show the average and standard deviation of performance metrics described in Section~\ref{sec_performance} obtained by our proposed metrics. Clusters based on $w_2$ group in relevant clusters the majority of the population in all the analysed PCs (reaching the maximum value of $0.94$ in PC1) to the detriment of precision, which falls to values between 0.22 and 0.35. As shown also in the previous investigation, the most promising similarity metrics in terms of precision and overlap ratio are both $w_7$ and $w_8$ followed by $w_5$. These outperform the other weights in all PCs, ensuring an overlap ratio within the same cluster with values in the range of $0.59$ and $0.70$ for $w_7$, $0.60$ and $0.72$ for $w_8$. Similarly, the values of precision are always over $0.42$ for both $w_7$ and $w_8$. The only exception is in PC1, where the best performing metric in terms of precision is $w_6$, which for the other contents cases is always the worst performing metric. 
\vspace{-1em}
\subsection{Trajectory-Based analysis}
\begin{figure*}[ht]
	\centering
    \hspace{-2em}
	\subfigure[Mean Overlap Ratio in Relevant Cluster.]{
	\includegraphics[width=0.35\textwidth]{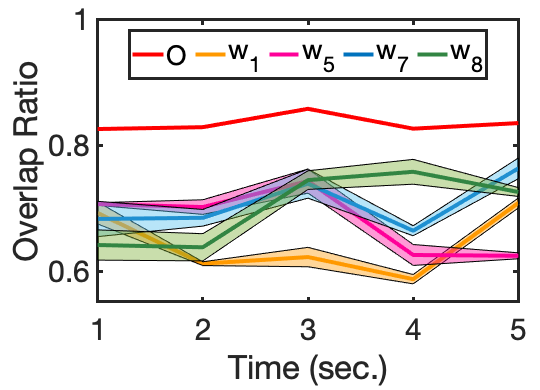}}
	\quad
    \hspace{-2em}
	\subfigure[Mean Relevant Users.]{
		\includegraphics[width=0.34\textwidth]{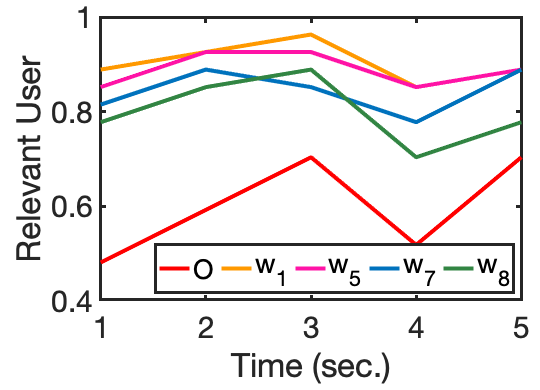}}
	\quad
    \hspace{-2em}
	\subfigure[Precision.]{
		\includegraphics[width=0.34\textwidth]{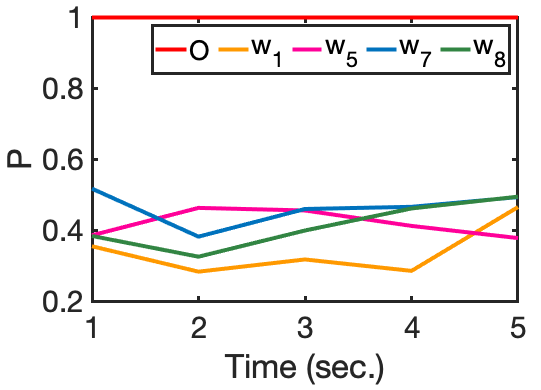}}
    \vspace{-1.5em}
	\caption{Clustering over time (chunk = 1 sec.) results per sequence PC1 (\textit{Longdress}): comparison between ground-truth $O$ and a subset of proposed metrics ($w_1$, $w_5$, $w_7$ and $w_8$).}
	\label{Fig_overTime_set2}
	\vspace{-1em}
\end{figure*}
Given the above remarks, we now analyse the performance metrics over time, taking into account only $w_1$, $w_5$, $w_7$, and $w_8$. Indeed, we decide to select the best-performing similarity metrics in the previous investigation ($w_5$, $w_7$ and $w_8$). To have a fair comparison, we also keep the most promising among the single-feature metrics, $w_1$. We compute clique-based clusters over a time window of $1s$ (\ie chunk) and a time similarity threshold of $0.8s$. At  each  chunk,  we  evaluate the average overlap ratio per relevant cluster, the average of the relevant population and the precision of detected clusters. As an example, we show  in Figure~\ref{Fig_overTime_set2} the performance results per sequence PC1 (\textit{Longdress}) as functions of time per each similarity metric. In Figure~\ref{Fig_overTime_set2}, we also add the performance of clusters detected by the ground-truth metric $O$ (\ie red line). The goal is indeed to find a metric able to perform similarly to our ground-truth over time. All the similarity metrics reach an average overlap ratio within clusters between $0.6$ and $0.75$ (Figure~\ref{Fig_overTime_set2}~(a)). However, clusters based on $w_1$ have lower performance, while other metrics are  performing quite similarly, although with a slight predominance of $w_7$. In terms of relevant users (Figure~\ref{Fig_overTime_set2}~(b)), it is worth noting that all the proposed similarity metrics generate bigger clusters than the ground-truth metric, which considers only half of the population as relevant. In more detail, the clusters resulting from $w_1$, $w_5$ and $w_8$ put in relevant clusters $0.8$ of the entire population for all the sequence time. Finally, in terms of precision as highlighted in Figure~\ref{Fig_overTime_set2}~(c) the only similarity metric that generated clusters with P over to $0.4$ in the entire sequence is $w_7$. These investigations show that similarity metrics based on multi-feature, such as $w_7$ and $w_8$, are more promising for detecting users with similar behaviour while experiencing volumetric content. 
\\
In summary, from this validation analysis, we can conclude the following: 
\begin{itemize}
    \item Overall,\textit{ multi-feature metrics} are more precise in detecting users with similar behaviour (in terms of displayed content) both in a frame- and chunk-based analysis;
    \item In particular, in spite of the slightly more complex formulation, $w_7$ and $w_8$ are robust and easy-to-use metrics that ensure a robust and reliable behavioural analysis via clustering tools;
    \item On the contrary, metrics based only on a single feature (\ie \textit{single-feature metrics}) are not sufficient to correctly identify similar users;
    \item The only exception among single-feature metrics is $w_1$ which is based only on the position of the user on the floor. Despite its simplicity, this metric is  comparable with multi-feature metrics. Hence, it can be used for an easy-to-implement preliminary behavioural analysis.
\end{itemize}
However, it is important to point out that these observations are currently only valid for similar volumetric contents (\ie human body). We leave further analysis across multiple datasets and types of content for future work.

\section{Discussion and Conclusion}
\label{sec_conclusion}
In this paper, we have summarised the main challenges of user behavioural analysis in a \sdof system due to the new settings and the added locomotion functionalities. Behavioural analysis of \sdof users is not considered in the literature yet; as such, there is no reference metric available to detect viewers who are displaying the same portion of the content. Thus, we considered a general ground-truth user similarity metric, such as \textit{overlap ratio}: the percentage of points displayed in common by two users. This is fairly straightforward, albeit time-consuming, to compute for point cloud contents, in which each point is rendered separately. For other types of volumetric contents, determining the overlap ratio is not as simple. Considering the number of vertexes that fall into a given frustum could lead to misleading results when large faces between sparsely distributed vertexes are present. Moreover, the metric requires rendering each volumetric video at any given time and for each viewer,  making its computation not trivial and intensely time-consuming. To overcome this issue and to assess users' similarity in a simple and objective way, we formulated and investigated several similarity metrics considering different distance features and measurements. We were interested in modelling similarities among users \textit{observing the same volumetric content}. In detail, we investigated different features or combinations of them which consider users' location in the virtual space and their viewing direction. We validated and tested our similarity metrics via a clique-based clustering tool proposed for \tdof scenario on real navigation trajectory collected in a \sdof VR environment. 
Therefore, in this article we advanced the state-of-the-art, proposing novel similarity metrics taking into account the new physical settings and locomotion functionalities given to users. Our results showed that solutions that consider both user position and viewing direction are promising  to correctly detect users with similar behaviour while experiencing volumetric content. Moreover, since these metrics are based on simple operations of data that are typically already known in a multimedia system (i.e., user position in the virtual space and viewing direction), they can be evaluated on average in a hundredth of a second. This makes our proposed metrics suitable for real-time applications. In future work, we will further test the robustness and versatility of these metrics on \sdof navigation trajectories collected in a different virtual scenario, for example in \ac{AR} applications \cite{zerman2021user}.





\bibliographystyle{ACM-Reference-Format}
\bibliography{refs}

\end{document}